\documentclass[12pt]{spieman}  
\usepackage{amsmath,amsfonts,amssymb}
\usepackage{graphicx}
\usepackage{setspace}
\usepackage{tocloft}
\usepackage{lineno}
\title{AI-Driven Segmentation and Analysis of Microbial Cells}

\author[a]{Shuang Zhang}
\author[b]{Carleton Coffin}
\author[c]{Karyn L. Rogers}
\author[b,*]{Catherine Ann Royer} 
\author[a,*]{Ge Wang}
\affil[a]{Department of Biomedical Engineering, Rensselaer Polytechnic Institute, USA 12180}
\affil[b]{Department of Biological Sciences, Rensselaer Polytechnic Institute, USA 12180}
\affil[c]{Department of Earth and Environmental Science, Rensselaer Polytechnic Institute, USA 12180}

\cftpagenumbersoff{Figure}
\cftpagenumbersoff{table} 
\begin{document} 
\maketitle

\begin{abstract}
Studying the growth and metabolism of microbes provides critical insights into their evolutionary adaptations to harsh environments, which are essential for microbial research and biotechnology applications. In this study, we developed an AI-driven image analysis system to efficiently segment individual cells and quantitatively analyze key cellular features. This system is comprised of four main modules. First, a denoising algorithm enhances contrast and suppresses noise while preserving fine cellular details. Second, the Segment Anything Model (SAM) enables accurate, zero-shot segmentation of cells without additional training. Third, post-processing is applied to refine segmentation results by removing over-segmented masks. Finally, quantitative analysis algorithms extract essential cellular features, including average intensity, length, width, and volume. The results show that denoising and post-processing significantly improved the segmentation accuracy of SAM in this new domain. Without human annotations, the AI-driven pipeline automatically and efficiently outlines cellular boundaries, indexes them, and calculates key cellular parameters with high accuracy. This framework will enable efficient and automated quantitative analysis of high-resolution fluorescence microscopy images to advance research into microbial adaptations to grow and metabolism that allow extremophiles to thrive in their harsh habitats.
\end{abstract}


{\noindent \footnotesize\textbf{*}Ge Wang,  \linkable{wangg6@rpi.edu}; Catherine Ann Royer, \linkable{royerc@rpi.edu} }

\begin{spacing}{2}   

\section{Introduction}
\label{sect:intro}  

A substantial portion of Earth’s microbial biomass resides in the extreme conditions such as high pressure, temperature, salinity, or pH \cite{kallmeyer2012global, inagaki2015exploring, trembath2017methyl, merino2019extremophiles, gallo2024advances}. They are key players in global biogeochemical cycles and serve as models for understanding the limits of life in planetary and astrobiological contexts\cite{merino2019extremophiles}, and offer insight into microbial evolution under high-pressure, early Earth–like conditions\cite{boyd2020phylogenomic}. Yet, their characterization at the single-cell level remains limited due to difficulties in cultivation, slow growth, and the lack of genetic tools for conventional reporter-based imaging \cite{rampelotto2024decade, riley2021approaches}. To overcome these barriers, researchers increasingly rely on intrinsic metabolic autofluorescence (e.g., NADH, FAD, F420) and recent advances in microscopy also enable high-resolution, in situ imaging and quantitative analysis of cells in their native environments \cite{stringari2012phasor, bourges2020quantitative}.

Among the key parameters accessible through imaging, cell morphology—particularly size and shape—serves as a sensitive indicator of microbial physiology and responses to environmental stress \cite{ishii2004ftsZ, young2006shape, campos2018size }. To extract this information from microscopy data, instance segmentation plays a central role by delineating individual cells with pixel-level precision. However, existing segmentation workflows often rely on manual annotation, which is time-consuming and limits scalability. To address this, there is a growing need for robust, automated segmentation methods that can accommodate the unique challenges of microbial imaging and support high-throughput, quantitative cell analysis in extremophiles and beyond.

Over the years, deep learning has dominated the field of cell segmentation. Encoder–decoder Convolutional Neural Network (CNN), particularly U-Net~\cite{ronneberger2015u} and its numerous variants, remain foundational architectures for microbial image segmentation due to their robustness, scalability, and ability to accurately delineate individual cells, even in complex environments. For example, Attention U-Net enhances the original architecture by integrating attention gates, allowing the model to focus on relevant cell regions and suppress background noise in microscopy images~\cite{oktayattention}. Instance segmentation methods like Mask R-CNN have been adapted from natural image analysis to distinguish individual cells, even in crowded settings, they detect cell proposals using a region proposal network and generate precise masks, making them useful for segmenting clustered or overlapping cells in fluorescence images~\cite{he2017mask}. Furthermore, transformer-based models such as Cell-DETR have been introduced, using detection transformers to directly predict cell instances via attention-based object queries, offering a powerful global context understanding~\cite{pina2024cell}. More recently, generalist models have gained traction due to their superior segmentation performance and generalizability across different imaging conditions.
Cellpose segments diverse cell types using a U-Net-like backbone that predicts spatial flow fields guiding pixels toward cell centers, enabling robust performance without retraining~\cite{stringer2021cellpose}. 
Its successor, Omnipose, refines this approach using distance field gradients, achieving high accuracy on irregularly shaped and densely packed bacterial cells~\cite{cutler2022omnipose}.

Despite these advancements, the above-described methods require building domain-specific segmentation datasets with many samples to train a segmentation model, which is expensive and time-consuming for a new cell imaging study.
The emergence of segmentation foundation models, such as the Segment Anything Model (SAM)~\cite{kirillov2023segment}, enables a paradigm without domain-specific dataset construction or additional training through the zero-shot learning ability.
Nevertheless, unique challenges such as noise interference, overlapping cell structures, blurry boundaries, and computational inefficiencies remain obstacles to achieving accurate segmentation with a foundation model.
To this end, we introduce a foundation model-driven cell segmentation pipeline that integrates denoising, segmentation, and post-processing for fast, accurate and quantitative analysis. Our approach enhances image quality using a denoising algorithm to suppress noise while preserving cellular details. It then employs SAM with appropriate prompting for zero-shot instance segmentation without requiring labeled training data. Finally, a post-processing pipeline refines segmentation masks and extracts meaningful quantitative parameters such as cell length, width, and volume. By integrating these techniques, our approach substantially improves segmentation accuracy, minimizes the need for manual intervention, enhances scalability for large-scale biological image analysis, and accelerates the overall processing pipeline. The following sections detail our methodology, present experimental results, and discuss the broader implications of our approach in biomedical research. 


\begin{figure}
\centering\includegraphics[width=1\textwidth]{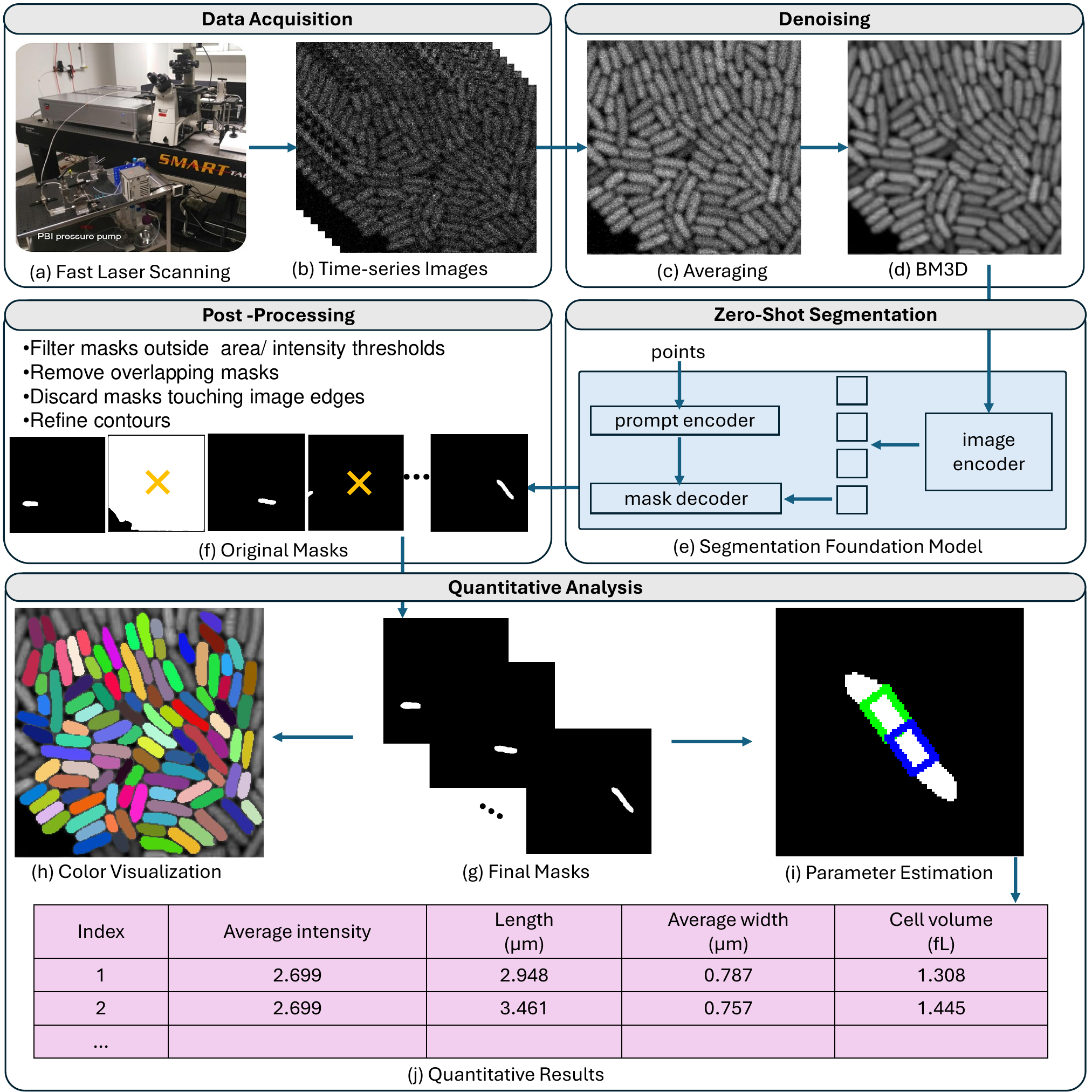}
\caption{\textbf{Foundation model-driven workflow for cell imaging.} \textbf{Data Acquisition:} A pressure and temperature regulated imaging system that utilizes 2 photon excitation and laser scanning microscopy (a) was used to generate time series images (b). \textbf{Denoising:} These images were averaged to produce a stacked image (c), enhancing signal-to-noise ratio. The stacked image was then denoised using the BM3D algorithm (d). \textbf{Zero-Shot Segmentation:} The denoised image was processed by the Segmentation Anything Model (SAM), a foundation model that generated initial cell masks using point prompts (e). \textbf{Post-processing:} Post-processing steps were applied to refine the masks (f). \textbf{Quantitative Analysis:} The final masks (g) were aggregated (h) and used for quantitative analysis (i), enabling the extraction of cellular features such as average intensity, length, average width, and cell volume (j).}
\label{fig_method}
\end{figure}

\section{Methodology}
Fig. \ref{fig_method} shows the AI-driven workflow for automated cell segmentation and quantitative morphological analysis. By running this AI-driven cell imaging system, we can automatically obtain detailed cellular features that are essential for studying microbial growth and metabolism, providing critical insights into their evolutionary adaptations to harsh environments. Detailed descriptions of each component of the workflow are provided in the following subsections.

\subsection{Data Acquisition}


As a proof of concept, the well-studied model organism \textit{Escherichia coli}, strain MG1655, was used to evaluate whether an AI-driven workflow could accurately segment rod-shaped microbes. Images were acquired using an ISS Alba fast scanning mirror fluctuation microscope, which employs two-photon excitation in combination with a high numerical aperture objective. Excitation was performed at 750~nm with approximately 40~mW of power to efficiently excite NADH and generate fluorescence images with well-defined cell boundaries, owing to the homogeneous intracellular distribution of NADH in \textit{E. coli}. This imaging setup also permits diffraction limited resolution, which enables precise morphological analysis.

To minimize photobleaching and phototoxicity—critical factors in quantitative fluorescence microscope techniques such as scanning number and brightness (sN\&B) and fluorescence lifetime imaging microscopy (FLIM), a pixel dwell time of 40~$\mu$s was used. A total of seven raster scans were collected per field of view (FOV), where the number of scans can be adjusted depending on the fluorescence imaging technique used. Each FOV comprised 256 $\times$ 256 pixels, corresponding to a physical area of 20~$\mu$m $\times$ 20~$\mu$m. This scale accommodates approximately 30 yeast cells or 100–300 bacterial cells per FOV, depending on cell size, shape, and species. Although imaging under extreme conditions such as high pressure presents additional challenges \cite{bourges2020quantitative}, the acquired data will always have the same file type (i.e. TIFF), making the AI-driven segmentation workflow applicable irrespective of hardware, software, or imaging conditions.

\subsection{Denoising}
Even though the fast scanning system is powerful, its speed leads to lower photon counts in each scan, which makes shot noise dominant at the lower limits of fluorescence detection. To facilitate the detection of true cellular features, the signal-to-noise ration (SNR) was improved by averaging the fluorescence intensity of the individual raster scans at each pixel as shown in Fig 1(c). Although the quality of the averaged image is improved compared to single raster scans, the noise remains significant and compromises the segmentation results. As such, we implemented a denoising algorithm to further refine image clarity, suppress residual background fluorescence fluctuations, and improve the accuracy of downstream segmentation and quantitative analysis. 

Specifically, we employed the Block-Matching and 3D Filtering (BM3D) algorithm~\cite{dabov2007image}, a method for image denoising that exploits self-similarity and sparsity in the transform domain. BM3D operates in two stages: the first stage performs block matching and grouping of similar image patches into 3D stacks, followed by collaborative filtering using a transform-domain thresholding process. In the second stage, a Wiener filter is applied in the transform domain to refine the estimates obtained in the first step.
In our implementation, we set the standard deviation to 0.2.

\subsection{Zero-Shot Segmentation}

Accurate instance segmentation is essential for extracting biological information. To maintain high accuracy while transitioning a traditionally manual process to an automated one, we employed zero-shot segmentation methods—without requiring extensive labeled data—including the Segment Anything Model (SAM) and its latest variant, SAM2, to explore their potential for achieving high-precision cell segmentation and to assess their applicability in subsequent analyses.

\subsubsection{Segment Anything Model (SAM)}

SAM is a foundation model designed for general-purpose segmentation. A key advantage of SAM is its zero-shot learning capability, enabling it to segment objects in previously unseen datasets without requiring task-specific fine-tuning. By leveraging large-scale pretraining, SAM generalizes well across a wide range of domains—including natural images and biomedical microscopy data—demonstrating strong versatility across different visual contexts~\cite{kirillov2023segment}. 

SAM operates in a prompt-based manner, where segmentation masks are generated based on user-provided prompts such as points, bounding boxes, or free-form text descriptions. The model consists of three primary components: an image encoder, a prompt encoder, and a mask decoder.
The image encoder, a vision transformer (ViT), takes a set of none-overlap patches as inputs and outputs their feature vectors to represent the input image. The prompt encoder processes user-defined prompts, which can be masks, points, boxes, or text, and outputs the prompt embedding vectors. Given the extracted image features and encoded prompts, the mask decoder will predict the segmentation masks and Intersection-over-Union (IoU) scores.

SAM is available in multiple variants, differing in model capacity and performance. The SAM-B (Base) uses the ViT-B backbone with approximately 91 million parameters, SAM-L (Large) uses ViT-L with 631 million parameters, and SAM-H (Huge) is built on ViT-H with around 2.4 billion parameters. Larger variants such as SAM-H offer better segmentation accuracy, especially for complex or noisy datasets, while requiring more computational resources. In microscopy imaging—where the accurate delineation of small, low-contrast, and densely packed structures is critical—larger model capacity significantly improves segmentation quality.









\subsubsection{Enhanced Segment Anything Model (SAM2)}

With the continuous development of foundation models, SAM2 was introduced as an enhanced version of SAM, incorporating several modifications aimed at improving segmentation efficiency and adaptability across different applications~\cite{ravi2024sam2}.
One key improvement in SAM2 is the refined prompt-handling mechanism. This enhancement allows the model to interpret segmentation cues more effectively, leading to more consistent and reliable mask predictions across various input conditions. Another major update is the optimization of mask selection. SAM2 improves the mask generation process by incorporating an attention-based filtering mechanism, ensuring that object delineation is more efficient and precise, particularly in complex scenes. Additionally, SAM2 introduces improved computational efficiency. The model has been optimized to maintain high segmentation accuracy while reducing inference time, making it more suitable for large-scale processing in biomedical applications. 

Like SAM, SAM2 is also released in multiple model variants based on ViT architecture, specifically SAM2-S (Small), SAM2-B (Base), and SAM2-L (Large). These correspond to ViT-S, ViT-B, and ViT-L backbones with approximately 46 million, 81 million, and 224 million parameters, respectively.










By evaluating both SAM and SAM2 within our segmentation pipeline, we aimed to assess their potential for achieving robust, adaptable, and high-precision segmentation in microscopy image analysis.In our implementation, we used a grid of $32 \times 32$ points as the prompts. 

\subsection{post-processing}

Post-processing is an essential step to refine segmentation results by eliminating artifacts, reducing redundancy, and enhancing accuracy. The raw segmentation outputs from SAM and SAM2 may contain small artifacts, duplicate masks, partial segmentations, or edge-touching objects, requiring additional processing before quantitative analysis. To address these issues, we apply a structured post-processing pipeline that filters, refines, and optimizes the segmented masks. The post-processing workflow consists of the following steps:

\emph{Initial Mask Filtering Based on Area and Intensity:}
 We set thresholds for area and intensity to remove small artifacts and irrelevant objects. Masks smaller than 100 pixels are discarded as noise, while those exceeding 1250 pixels are removed as they may correspond to merged objects or background artifacts. Additionally, we apply intensity filtering to eliminate artifacts caused by incorrect segmentation. The mean intensity of all selected cells (\(\bar{I}\)) is computed, and masks deviating significantly from this mean, defined as \( 0.35\bar{I} < I_i < 1.6\bar{I} \), are removed. These thresholds are selected through preliminary experimentation.

\emph{Removing Fully Contained Masks:}When using SAM, which tends to over-segment objects by treating distinct parts as separate instances. For example, in natural images, SAM may segment a car into multiple overlapping regions such as the body, windows, and tires. Similarly, in microscopy images, a single cell can be incorrectly segmented into one large mask along with one or two smaller interior masks. To mitigate this, we discard smaller masks that are completely contained within larger ones. Formally, given two masks \( M_i \) and \( M_j \), if \( M_i \subseteq M_j \), the smaller mask is removed to prevent duplicate segmentations of the same object from being counted multiple times. 

\emph{Non-Maximum Suppression (NMS) for Overlapping Masks:} To address over-segmentation caused by closely overlapping masks, we implement Non-Maximum Suppression (NMS) based on pairwise Intersection over Union (IoU). For each mask \( M_i \), we compute the IoU with all other masks \( M_j \), defined as:
\begin{equation}
    \text{IoU}(M_i, M_j) = \frac{|M_i \cap M_j|}{|M_i \cup M_j|}.
\end{equation}
If the IoU between \( M_i \) and \( M_j \) exceeds a threshold (set to 0.3 in this study), the mask with the smaller area is removed. This approach ensures that among highly overlapping masks, only the largest and most representative one is retained, thereby preventing redundant segmentation of the same object. 

\emph{Erosion-Based Partial Overlap Removal:}
To further refine overlapping masks, we apply morphological erosion, which shrinks the smaller mask before checking its containment within another mask. The eroded mask is given by
\begin{equation}
    M'_i = M_i \ominus K,
\end{equation}
where \( K \) is a structuring element. If the eroded mask remains fully enclosed within another segmentation, it is removed, preventing multiple segmentations of closely overlapping objects.

\emph{Edge Mask Removal:}
Masks that intersect image boundaries are often incomplete and unreliable. To ensure that only fully visible cells are retained, we remove any mask \( M_i \) that has pixels within the first or last two rows/columns of the image. The updated condition is:
\begin{equation}
    \exists (x, y) \in M_i, \quad x \leq 1 \text{ or } x \geq W - 2 \text{ or } y \leq 1 \text{ or } y \geq H - 2.
\end{equation}
This adjustment ensures that cells close to the boundary, which may be partially segmented within the FOV, are excluded from further analysis.

\emph{Morphological Closing for Smoother Contours:}
To refine the final segmentation results, we apply morphological closing, which consists of dilation followed by erosion. This process smooths jagged edges and fills small gaps in the masks, enhancing the structural integrity of segmented objects. The updated mask is computed as:
\begin{equation}
    M' = (M \oplus K) \ominus K.
\end{equation}
where \( \oplus \) represents dilation and \( \ominus \) represents erosion with structuring element \( K \). This operation ensures a more precise cell shape representation. By integrating this post-processing pipeline, we achieve more reliable segmentation results, better-defined cell boundaries, reduced redundant masks, and improved accuracy in quantitative cell analysis.

To evaluate the accuracy of our method, we define an error rate based on manual annotation. We manually mark the locations of cells and then count the number of incorrect detections in the segmentation results, including wrong position and missed cells. The error rate is calculated as the ratio of incorrect detections to the total number of cells.

\subsection{Quantitative Analysis}
Quantitative analysis algorithms were employed to extract key cellular features, including average intensity, length, average width, and volume for each cell. Average intensity was calculated by overlaying the cell mask onto the stacked image and computing the mean pixel value within the mask. To estimate cell volume, we constructed a geometric model for each cell, assuming its cross-section consists of two semicircles and a rectangle—approximated in 3D as two hemispheres and a cylinder.

First, a rectangular bounding box with maximum overlap was fitted to each cell mask to estimate the cell length \( L \). To determine the average width \( W \), the bounding box was divided into four equal sections along its length. The middle two sections, refined to better represent the cell body, were used to compute the average width. Using these measurements, the cell volume \( V \) was then calculated according to the geometric model as:
\[
V = \pi \left( \frac{W}{2} \right)^2 (L-W) + \frac{4}{3} \pi \left( \frac{W}{2} \right)^3.
\]
These extracted morphological features enable the precise characterization of cellular structure and provide valuable insights into cellular adaptation.

\section{Results}
\subsection{Denoising}
\begin{figure}[htbp]
\centering\includegraphics[width=13cm]{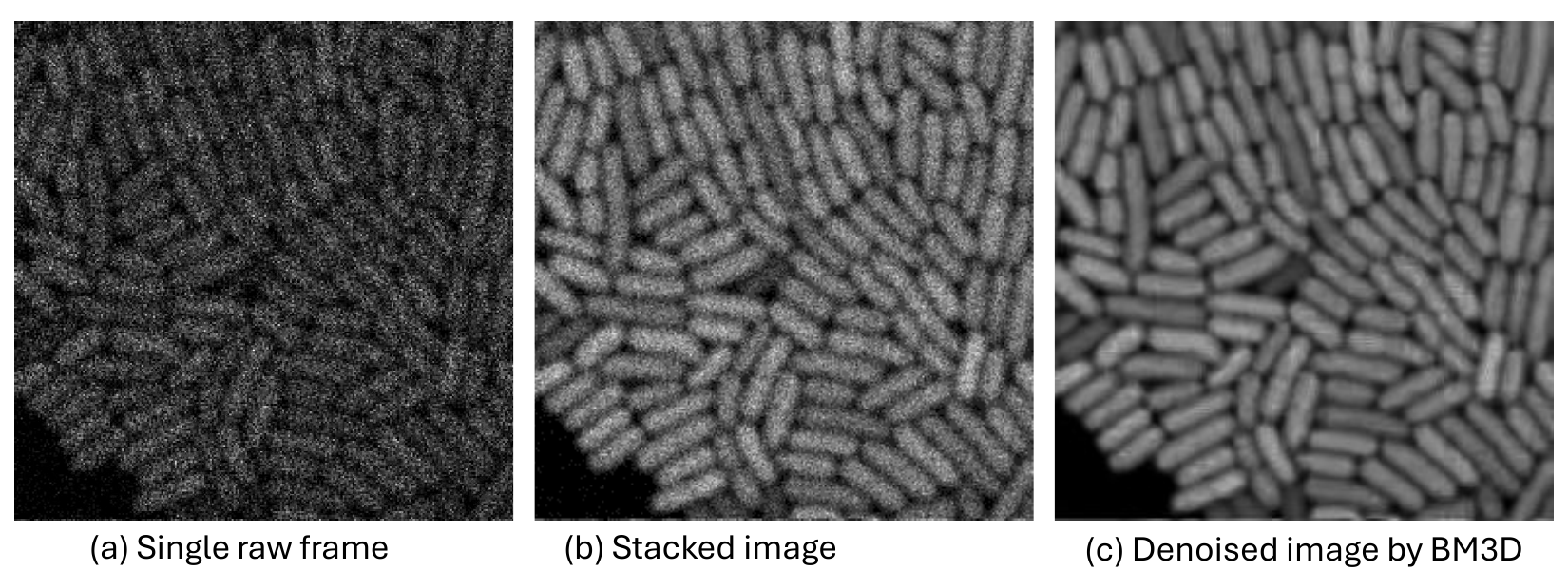}
\caption{Comparison of raw, stacked, and BM3D-denoised cell images.}
\label{fig_denoising}
\end{figure}

Fig. \ref{fig_denoising} shows how time-series image stacking and BM3D denoising enhance the quality of noisy cell images. Fig. \ref{fig_denoising}(a) represents a single-frame raw image, which exhibits a high level of noise due to photon shot and dark noise, making individual cells difficult to distinguish. Fig. \ref{fig_denoising}(b) shows the stacked image, generated by averaging multiple frames, which effectively reduces noise by leveraging temporal redundancy. However, while this approach improves visibility, residual noise remains, and cell boundaries are still blurred. 
Fig.~\ref{fig_denoising}(c) shows the BM3D-denoised image, which effectively reduces noise while preserving fine cellular structures and boundary integrity. This results in improved contrast which will enhance the segmentation accuracy and downstream quantitative analysis.

These results highlight the effectiveness of BM3D in microscopy image preprocessing, ensuring high-fidelity cell morphology representation while minimizing noise-related artifacts. The improved signal-to-noise ratio (SNR) achieved through BM3D denoising is particularly beneficial for robust cell segmentation and feature extraction in subsequent processing steps.

\subsection{Segmentation Results}
\begin{figure}[h]
\centering\includegraphics[width=13cm]{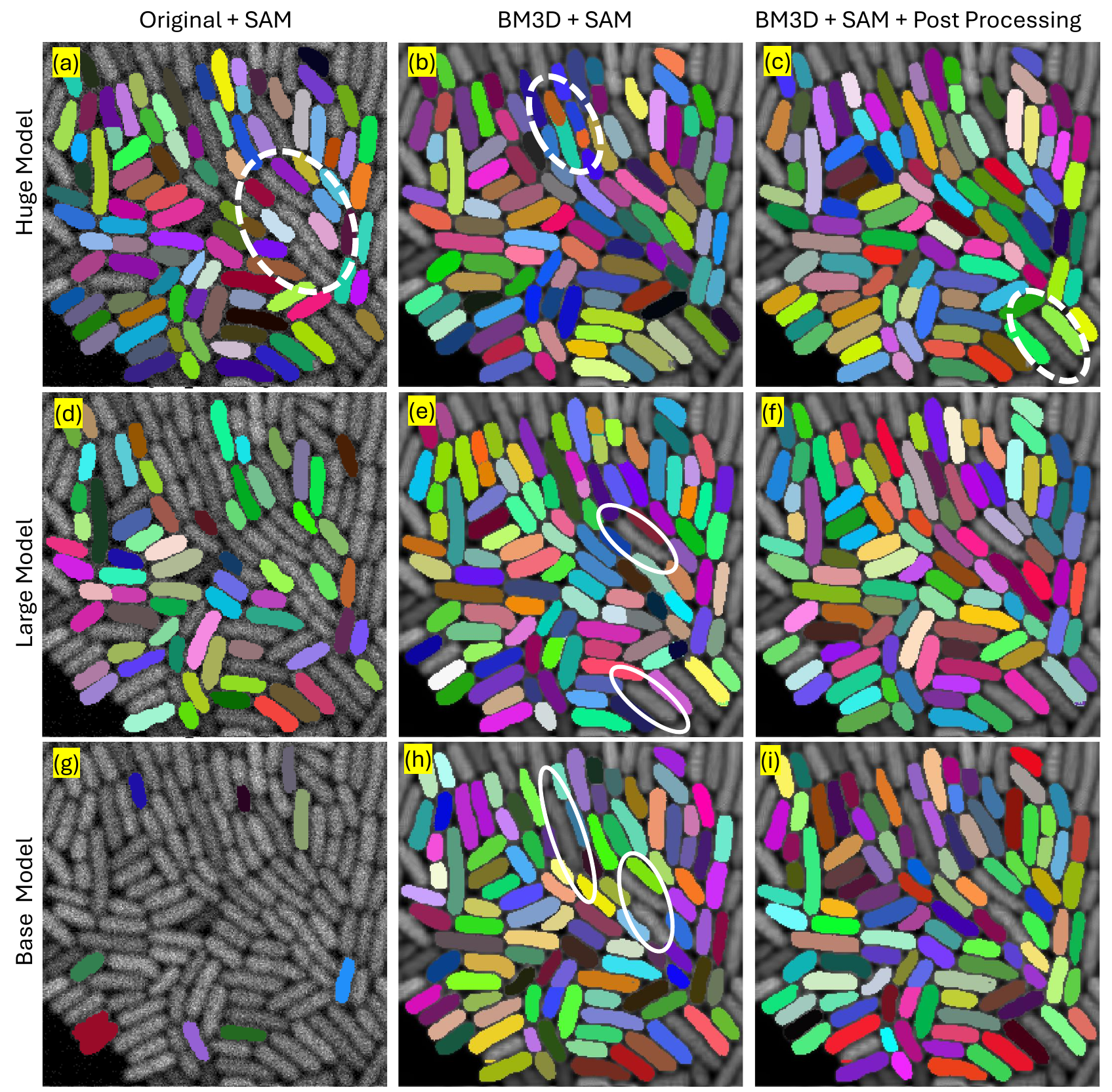}
\caption{Effect of BM3D denoising and post-processing on SAM.The first column (a, d, g) displays segmentation results on raw microscopy images, which are characterized by high noise levels and poor contrast. The second column (b, e, h) presents outputs following BM3D denoising, while the third column (c, f, i) shows the results after an additional post-processing step.}
\label{fig_SAM}
\end{figure}

\begin{figure}[h]
\centering\includegraphics[width=13cm]{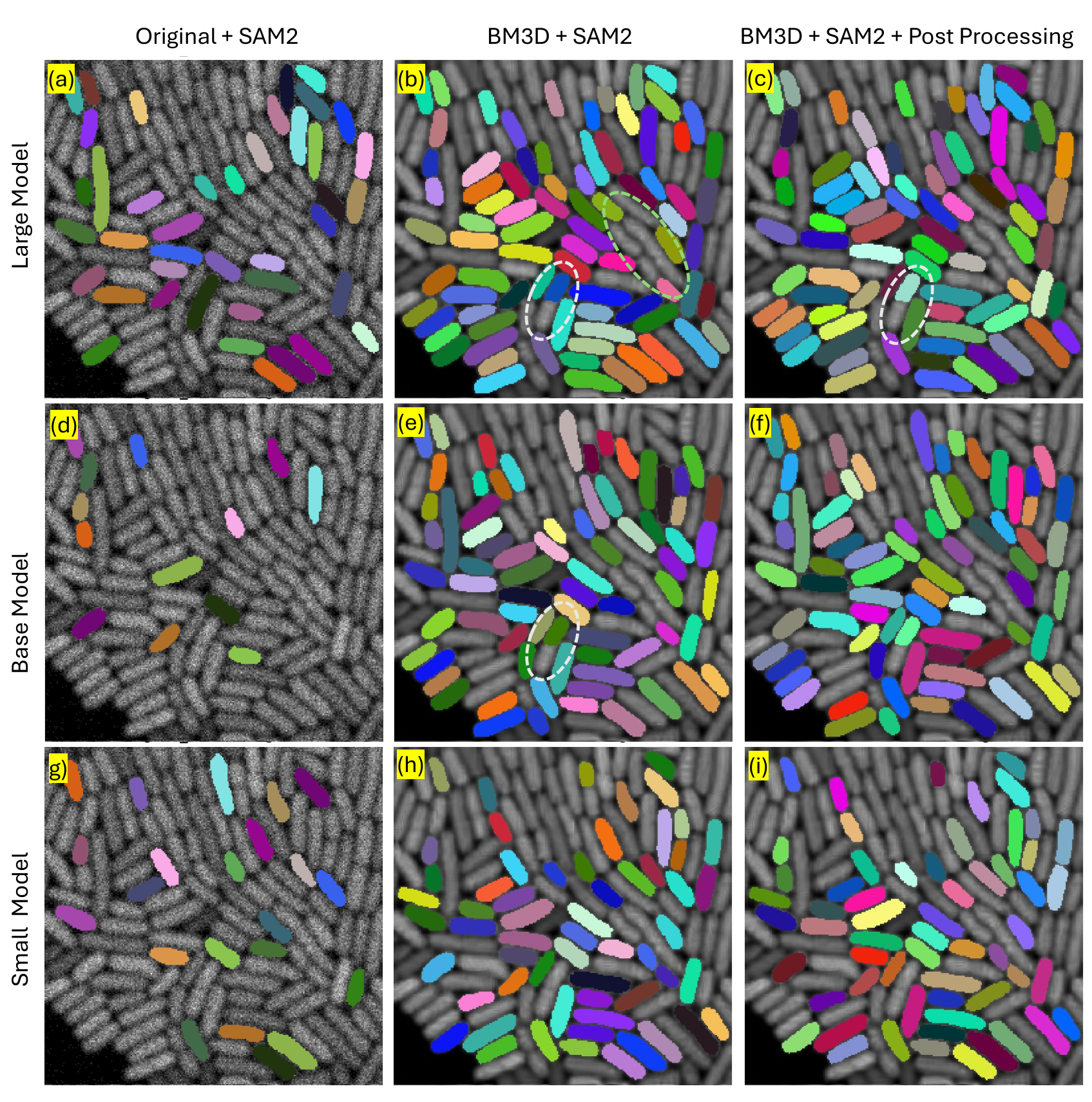}
\caption{Effect of BM3D denoising and post-processing on SAM2. The first column (a, d, g) displays segmentation results on raw microscopy images, which are characterized by high noise levels and poor contrast. The second column (b, e, h) presents outputs following BM3D denoising, while the third column (c, f, i) shows the results after an additional post-processing step.}
\label{fig_SAM2}
\end{figure}

Fig. \ref{fig_SAM} illustrates the segmentation performance of three SAM variants—SAM-B (Base), SAM-L (Large), and SAM-H (Huge)—under different preprocessing configurations. When segmenting raw images, clear differences emerge among the SAM variants. SAM-H is able to detect a significantly higher number of cells compared to SAM-L and SAM-B, suggesting that the larger vision transformer architecture enhances the model's capacity to generalize and adapt to challenging object detection tasks—even under suboptimal image quality. However, certain cells remain undetected when using a noisy image as input, as indicated by the white circle in Fig. \ref{fig_SAM}(a), highlighting the limitations imposed by extreme noise even for the most capable variant.
The application of BM3D denoising leads to a substantial improvement in segmentation quality, particularly for SAM-L and SAM-B. Enhanced contrast and reduced background noise allow these models to better resolve cell boundaries and produce a greater number of valid segmentation masks. These observations underscore the importance of input quality: with sufficient denoising, all SAM variants demonstrate notable performance improvements, with SAM-L maintaining a modest advantage in segmentation accuracy. Across all tested conditions, SAM-H consistently yields the most accurate and complete segmentation. In contrast, SAM-B and SAM-L frequently fail to capture finer morphological details, as evidenced by the white ellipses in Figs. \ref{fig_SAM}(e) and (h). However, SAM-H is not without limitations. As shown in Fig. \ref{fig_SAM2}(b), it occasionally exhibits over-segmentation—splitting a single cell into multiple masks—as illustrated by the white-circled examples, where the closely adjacent cells are both segmented as two overlapped objects.

Post-processing results, shown in Fig. \ref{fig_SAM}(c), demonstrate the effectiveness of filtering steps in eliminating small, overlapping, or low-quality masks. This refinement enhances segmentation precision and yields a cleaner representation of the cell population. Although one cell remains undetected in Fig.~\ref{fig_SAM}(c), the overall missed detection rate is as low as 0.9\%, demonstrating strong segmentation performance.


In summary, this evaluation demonstrates that high-quality input images and carefully designed post-processing are critical for achieving accurate cell segmentation. While larger vision transformers like SAM-H offer superior robustness and detection capacity, preprocessing and prompt localization continue to play essential roles in maximizing segmentation accuracy and consistency.

Fig. \ref{fig_SAM2} presents the segmentation performance of three SAM2 variants—SAM2-S (Small), SAM2-B (Base), and SAM2-L (Large)—under the same preprocessing conditions. From Fig. \ref{fig_SAM2}(b,e,h), we observe that SAM2 yields cleaner mask boundaries with minimal overlap, which may benefit from architectural improvements. However, a limitation is evident in Fig. \ref{fig_SAM2}(e, h), where some cells (highlighted with white dotted lines) are segmented only partially—capturing roughly half of the cell body—indicating that SAM2 may occasionally fail to resolve entire objects, particularly under challenging imaging conditions. 

When compared to the original SAM models (Fig. \ref{fig_SAM}), SAM2 demonstrates a generally lower segmentation sensitivity in low-contrast microscopy images, with this deficiency most notable in the SAM2-L variant. Despite its improvements in boundary clarity, SAM2 fails to identify numerous cells—especially in the denoised and post-processed outputs—resulting in reduced overall mask coverage. This under-segmentation is particularly pronounced in densely packed cellular regions, as highlighted by the green-circled area in Fig. \ref{fig_SAM}(b). 
These performance limitations may stem from the relatively smaller model size and reduced representation capacity of SAM2 compared to the original SAM models. While SAM2 is optimized, such architectural modifications may compromise its ability to generalize in challenging biomedical imaging contexts. For cell segmentation tasks—particularly those requiring high fidelity in dense, low-contrast environments—the original SAM, especially its larger variants, appears more suitable due to its superior capacity to preserve fine structural details and achieve comprehensive mask coverage. This behavior aligns with recent findings by Sengupta et al.~\cite{sengupta2024is} who evaluated SAM and SAM2 in multiple medical imaging modalities. Their results indicate that SAM2 does not consistently outperform SAM, particularly in low-contrast biomedical images. These limitations are consistent with the challenges observed in our microscopy data, where precise delineation of small and densely packed cells is critical. 


For the task of cell segmentation in low-contrast microscopy images, the original SAM-H model demonstrates superior robustness and accuracy, making it the most suitable choice among the evaluated models. These results indicate that our AI system is well-positioned to integrate more advanced segmentation foundation models as they become available, enabling further improvements in biomedical image analysis.


\subsection{Post-processing}

\begin{figure}[h]
\centering\includegraphics[width=16cm]{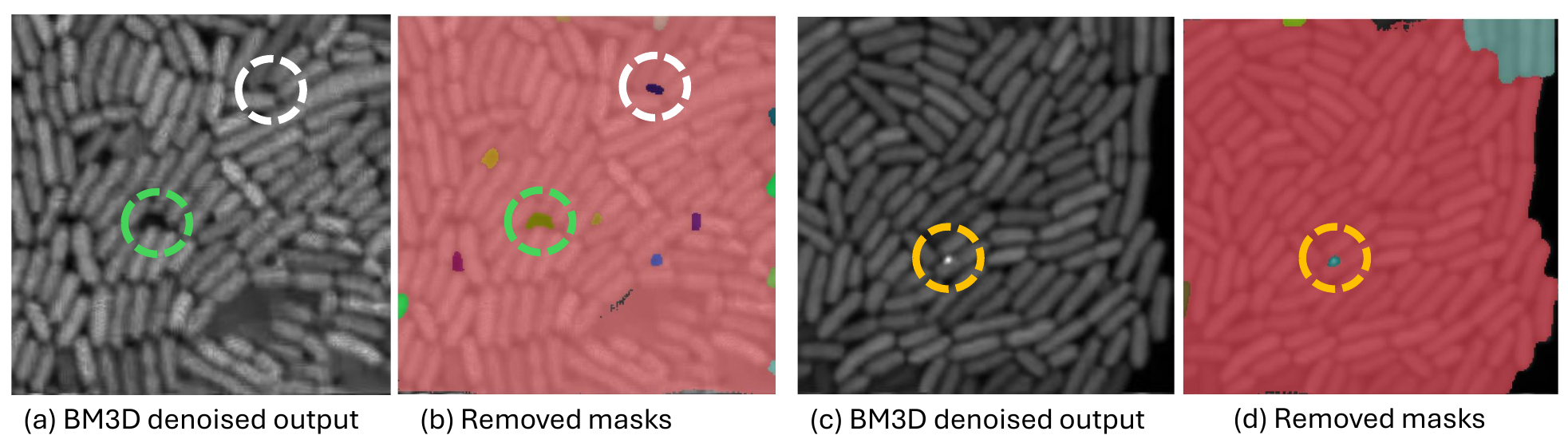}
\caption{Examples of cell masks removed due to abnormal area or intensity. (a) and (c) show the denoised input images, while (b) and (d) highlight the removed masks overlaid on the denoised image. The large pink regions, white-circled mask, green-circled mask, and yellow-circled mask correspond to masks excluded due to extreme values: large area, small area, low pixel intensity, and high pixel intensity, respectively. }
\label{fig_filtered}
\end{figure}

To demonstrate how the post-processing pipeline affects the segmentation results, we present representative filtered masks in separate images. Fig.~\ref{fig_filtered} illustrates examples of masks that were removed based on abnormal area or intensity criteria. Fig.~\ref{fig_filtered}(a) and (c) show two different denoised cell images, which are used to indicate the original locations of the removed segmented masks shown in Fig.~\ref{fig_filtered}(b) and (d), respectively. Specifically, the large pink masks represent over-segmented regions that cover nearly the entire image and are excluded due to their excessively large area. The masks outlined in white and green in Fig.~\ref{fig_filtered}(a) correspond to empty spaces between cells. The white-circled mask can be removed by both the small-area and low-intensity filters, while the green-circled mask is excluded due to low intensity. Additionally, the yellow-circled mask, due to its high intensity, is also removed based on predefined criteria. These examples underscore the importance of combining intensity- and size-based filtering to improve segmentation quality by eliminating outliers and retaining only biologically meaningful cell masks.

\begin{figure}[htbp]
\centering\includegraphics[width=13cm]{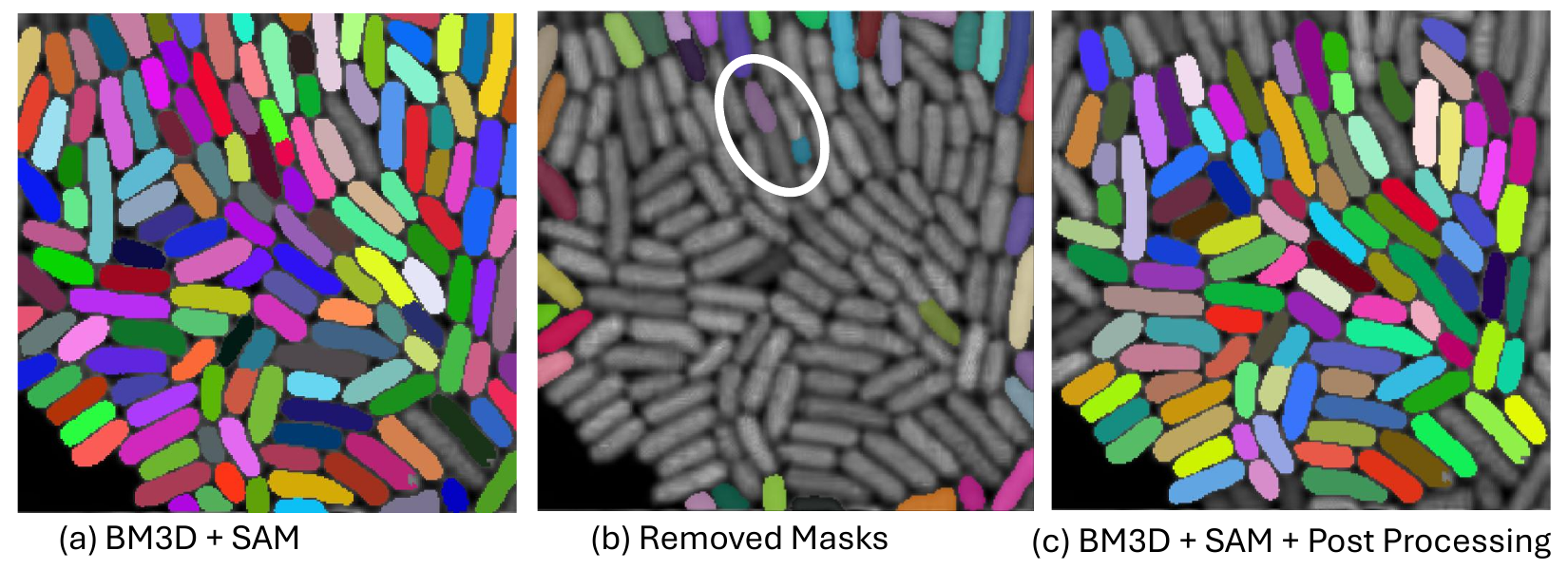}
\caption{Examples of masks removed due to edge contact or overlapping regions. (a) shows the initial segmentation results obtained using BM3D denoising followed by SAM. (b) displays the removed masks overlaid on the denoised image. Masks near the image boundary were excluded to avoid partial cell artifacts, while the white-circled masks represent over-segmented cells. (c) presents the final segmentation result after post-processing, with refined masks that exclude overlapping or edge-touching regions.}
\label{fig_filtered2}
\end{figure}

Fig. \ref{fig_filtered2}(b) illustrates masks that were removed due to their proximity to image boundaries or significant overlap with other masks. Specifically, cell masks that are truncated by the image edge are excluded to prevent incomplete or inaccurate measurements. By applying this step, the segmentation pipeline ensures that only well-contained and non-overlapping masks are retained, leading to more reliable downstream quantitative analysis.

To quantitatively assess segmentation accuracy across different configurations, we compute the error rate by comparing predicted masks to manually annotated ground truth. A total of 1,162 cells were manually labeled across 10 microscopy images. The resulting accuracy metrics are summarized in Table~\ref{tab:error_rate}. Among all evaluated settings, the combination of BM3D denoising and post-processing consistently yields the lowest error rate. In particular, SAM-H with predenoising and postprocessing achieves an average error rate of just 3.0\%, significantly outperforming all other variants. In contrast, SAM2-L exhibits a substantially higher error rate of 16.0\%. These findings validate the effectiveness of our proposed post-processing pipeline and further emphasize the superior segmentation precision of SAM-H, especially in dense, low-contrast cellular environments.

\begin{table}[ht]
\caption{Error rate comparison between SAM-H and SAM2-L under different processing stages.}
\label{tab:error_rate}
\begin{center}
\begin{tabular}{|c|c|c|c|c|c|c|c|}
\hline
\multicolumn{2}{|c|}{\rule[-1ex]{0pt}{3.5ex}\textbf{Error Rate \%}} &
\multicolumn{3}{c|}{\textbf{SAM-H}} & 
\multicolumn{3}{c|}{\textbf{SAM2-L}} \\
\hline
\rule[-1ex]{0pt}{3.5ex} \textbf{Image} &
\textbf{\# of Cells} &
\shortstack{Original\\+ SAM} & 
\shortstack{BM3D\\+ SAM} & 
\shortstack{BM3D \\+ SAM+ PP} & 
\shortstack{Original\\+ SAM} & 
\shortstack{BM3D\\+ SAM} & 
\shortstack{BM3D +\\ SAM+ PP} \\
\hline\hline
\rule[-1ex]{0pt}{3.5ex} 1 & 107 & 13.10 & 6.50 & 2.80 & 87.90 & 31.80 & 31.80 \\
\hline
\rule[-1ex]{0pt}{3.5ex} 2 & 105 & 2.90 & 4.80 & 1.00 & 84.80 & 10.50 & 10.50 \\
\hline
\rule[-1ex]{0pt}{3.5ex} 3 & 99  & 9.10 & 6.10 & 3.00 & 73.70 & 17.20 & 17.20 \\
\hline
\rule[-1ex]{0pt}{3.5ex} 4 & 112 & 6.30 & 8.00 & 4.50 & 83.90 & 27.70 & 27.70 \\
\hline
\rule[-1ex]{0pt}{3.5ex} 5 & 105 & 61.90 & 12.40 & 7.60 & 95.70 & 30.40 & 30.40 \\
\hline
\rule[-1ex]{0pt}{3.5ex} 6 & 95  & 36.80 & 4.20 & 4.20 & 96.80 & 20.00 & 20.00 \\
\hline
\rule[-1ex]{0pt}{3.5ex} 7 & 98  & 36.70 & 7.10 & 4.10 & 92.90 & 14.30 & 14.30 \\
\hline
\rule[-1ex]{0pt}{3.5ex} 8 & 142 & 2.80 & 1.40 & 1.40 & 78.20 & 2.10 & 2.10 \\
\hline
\rule[-1ex]{0pt}{3.5ex} 9 & 143 & 3.50 & 0.70 & 0.70 & 83.20 & 0.00 & 0.00 \\
\hline
\rule[-1ex]{0pt}{3.5ex} 10 & 156 & 0.60 & 1.30 & 0.60 & 86.50 & 6.40 & 6.40 \\
\hline
\multicolumn{2}{|c|}{\rule[-1ex]{0pt}{3.5ex}\textbf{Average}} & \textbf{17.40} & \textbf{5.30} & \textbf{3.00} & \textbf{86.40} & \textbf{16.00} & \textbf{16.00} \\
\hline
\end{tabular}
\end{center}
\end{table}

\section{Quantitative Features}
\begin{figure}[h]
\centering\includegraphics[width=9cm]{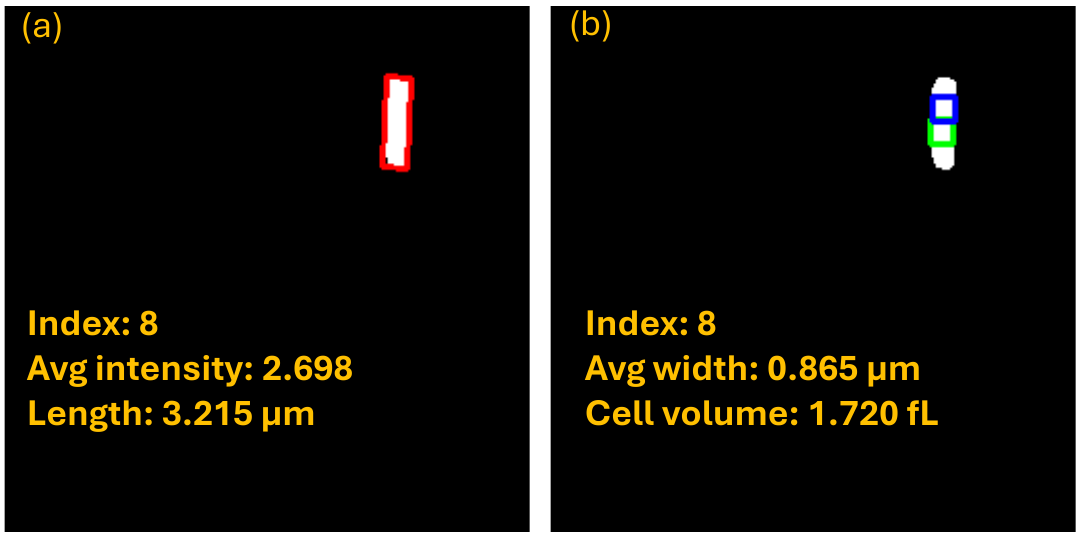}
\caption{Illustration of quantitative feature extraction for a single cell. (a) The cell length was computed from the major axis of the red bounding box surrounding the cell, while the average intensity was calculated by overlaying the cell mask onto the stacked image, resulting in a length of 3.215~\textmu m and an average intensity of 2.698. (b) The average width 0.865~\textmu m was estimated from the central two bounding rectangles, and the cell volume 1.720~fL was computed using a geometric model consisting of a cylinder and two hemispherical caps.}
\label{fig_cellsize}
\end{figure}

Fig. \ref{fig_cellsize} illustrates the quantitative feature extraction for a single cell. The average intensity and length were computed by overlaying the cell mask onto the stacked image, as shown in Fig. \ref{fig_cellsize}(a), resulting in an average intensity of 2.698 and a length of 3.215~\textmu m. The average width was measured from the central two rectangles, and the cell volume was estimated based on a geometric model consisting of a cylinder and two hemispheres. As shown in Fig. \ref{fig_cellsize}(b), the average width was 0.865~\textmu m,and the calculated cell volume was 1.720~fL.

\section{Conclusion}

In this study, we introduced an AI-driven pipeline for high-resolution microscopy image segmentation and analysis, incorporating BM3D-based denoising, the Segment Anything Model (SAM), a structured post-processing routine, and a quantitative analysis workflow. This integrated approach effectively addresses key challenges in biological imaging, including noise, low contrast, and overlapping cell structures. Quantitative evaluations show that the SAM-Huge variant, when paired with denoising and post-processing, achieves the highest segmentation accuracy, with an average error rate of just 3.0\%. In contrast, SAM2-Large performs less effectively in resolving fine structural details under low-contrast conditions. Beyond segmentation, our pipeline enables precise extraction of essential morphological features—including intensity, length, width, and volume—through a geometric modeling approach. These capabilities would facilitate in-depth studies of microbial metabolism, structural adaptation, and evolutionary dynamics in extreme environments. Overall, the proposed framework provides a scalable, label-free, and high-fidelity solution for quantitative cellular analysis, with strong potential for extension to a wide range of biomedical imaging applications requiring accurate structural characterization.

\subsection* {Acknowledgments}
This research is supported by the DOE Office of Science, Office of Biological and Environmental Research (BER), grant no. 280597.


\bibliography{report}   

\begin{thebibliography}{10}

\bibitem{kallmeyer2012global}
J.~Kallmeyer, R.~Pockalny, R.~R. Adhikari, {\em et~al.}, ``Global distribution of microbial abundance and biomass in subseafloor sediment,'' {\em Proceedings of the National Academy of Sciences} {\bf 109}(40), 16213--16216  (2012).
\newblock [doi:10.1073/pnas.1203849109].

\bibitem{inagaki2015exploring}
F.~Inagaki, K.-U. Hinrichs, Y.~Kubo, {\em et~al.}, ``Exploring deep microbial life in coal-bearing sediment down to ~2.5 km below the ocean floor,'' {\em Science} {\bf 349}(6246), 420--424  (2015).
\newblock [doi: 10.1126/science.aaa688].

\bibitem{trembath2017methyl}
E.~Trembath-Reichert, Y.~Morono, A.~Ijiri, {\em et~al.}, ``Methyl-compound use and slow growth characterize microbial life in 2-km-deep subseafloor coal and shale beds,'' {\em Proceedings of the National Academy of Sciences} {\bf 114}(44), E9206--E9215  (2017).
\newblock [doi: 10.1073/pnas.1707525114].

\bibitem{merino2019extremophiles}
N.~Merino, H.~S. Aronson, D.~P. Bojanova, {\em et~al.}, ``Living at the extremes: Extremophiles and the limits of life in a planetary context,'' {\em Frontiers in Microbiology} {\bf 10}, 780  (2019).
\newblock [doi: 10.3389/fmicb.2019.00780].

\bibitem{gallo2024advances}
G.~Gallo and M.~Aulitto, ``Advances in extremophile research: Biotechnological applications through isolation and identification techniques,'' {\em Life} {\bf 14}(9), 1205  (2024).
\newblock [doi: 10.3390/life14091205].

\bibitem{boyd2020phylogenomic}
D.~R. Colman, M.~R. Lindsay, M.~J. Amenabar, {\em et~al.}, ``Phylogenomic analysis of novel diaforarchaea is consistent with sulfite but not sulfate reduction in volcanic environments on early earth,'' {\em The ISME Journal} {\bf 14}(5), 1316--1331  (2020).
\newblock [doi: 10.1038/s41396-020-0611-9].

\bibitem{rampelotto2024decade}
P.~H. Rampelotto, ``Extremophiles and extreme environments: A decade of progress and challenges,'' {\em Life} {\bf 14}(3), 382  (2024).
\newblock [doi: 10.3390/life14030382].

\bibitem{riley2021approaches}
L.~A. Riley and A.~M. Guss, ``Approaches to genetic tool development for rapid domestication of non-model microorganisms,'' {\em Biotechnology Biofuels} {\bf 14}, 30  (2021).
\newblock [doi: 10.1186/s13068-020-01872-z].

\bibitem{stringari2012phasor}
C.~Stringari, A.~Cinquin, O.~Cinquin, {\em et~al.}, ``Phasor approach to fluorescence lifetime microscopy distinguishes different metabolic states of germ cells in a live tissue,'' {\em Proceedings of the National Academy of Sciences} {\bf 108}(33), 13582--13587  (2011).
\newblock [doi: 10.1073/pnas.1108161108].

\bibitem{bourges2020quantitative}
A.~C. Bourges, A.~Lazarev, N.~Declerck, {\em et~al.}, ``Quantitative high-resolution imaging of live microbial cells at high hydrostatic pressure,'' {\em Biophysical Journal} {\bf 118}(11), 2670--2679  (2020).
\newblock [doi: 10.1016/j.bpj.2020.04.017].

\bibitem{ishii2004ftsZ}
A.~Ishii, T.~Sato, M.~Wachi, {\em et~al.}, ``Effects of high hydrostatic pressure on bacterial cytoskeleton ftsz polymers in vivo and in vitro,'' {\em Microbiology} {\bf 150}(6), 1965--1972  (2004).
\newblock [doi: 10.1099/mic.0.26962-0].

\bibitem{young2006shape}
K.~D. Young, ``The selective value of bacterial shape,'' {\em Microbiology and Molecular Biology Reviews} {\bf 70}(3), 660--703  (2006).
\newblock [doi: 10.1128/MMBR.00001-06].

\bibitem{campos2018size}
M.~Campos, I.~V. Surovtsev, S.~Kato, {\em et~al.}, ``A constant size extension drives bacterial cell size homeostasis,'' {\em Cell} {\bf 159}(6), 1433--1446  (2014).
\newblock [doi: 10.1016/j.cell.2014.11.022].

\bibitem{ronneberger2015u}
O.~Ronneberger, P.~Fischer, and T.~Brox, ``U-net: Convolutional networks for biomedical image segmentation,'' in {\em Medical Image Computing and Computer-Assisted Intervention -- MICCAI 2015},  N.~Navab, J.~Hornegger, W.~M. Wells, {\em et~al.}, Eds., 234--241, Springer International Publishing, (Cham)  (2015).

\bibitem{oktayattention}
O.~Oktay, J.~Schlemper, L.~Le~Folgoc, {\em et~al.}, ``Attention u-net: Learning where to look for the pancreas,'' {\em ArXiv}   (2018).
\newblock [doi: 10.48550/arXiv.1804.03999].

\bibitem{he2017mask}
K.~He, G.~Gkioxari, P.~Dollár, {\em et~al.}, ``Mask r-cnn,'' in {\em 2017 IEEE International Conference on Computer Vision (ICCV)},  2980--2988  (2017).
\newblock [doi: 10.1109/ICCV.2017.322].

\bibitem{pina2024cell}
O.~Pina, E.~Dorca, and V.~Vilaplana, ``Cell-detr: Efficient cell detection and classification in wsis with transformers,'' in {\em Proceedings of The 7nd International Conference on Medical Imaging with Deep Learning},  N.~Burgos, C.~Petitjean, M.~Vakalopoulou, {\em et~al.}, Eds., {\em Proceedings of Machine Learning Research} {\bf 250}, 1128--1141, PMLR  (2024).

\bibitem{stringer2021cellpose}
C.~Stringer, T.~Wang, M.~Michaelos, {\em et~al.}, ``Cellpose: a generalist algorithm for cellular segmentation,'' {\em Nature methods} {\bf 18}, 100--106  (2021).
\newblock [doi: 10.1038/s41592-020-01018-x].

\bibitem{cutler2022omnipose}
K.~J. Cutler, C.~Stringer, T.~W. Lo, {\em et~al.}, ``Omnipose: a high-precision morphology-independent solution for bacterial cell segmentation,'' {\em Nature methods} {\bf 19}, 1438--1448  (2022).
\newblock [doi: 10.1038/s41592-022-01639-4].

\bibitem{kirillov2023segment}
A.~Kirillov, E.~Mintun, N.~Ravi, {\em et~al.}, ``Segment anything,'' {\em arXiv:2304.02643}   (2023).
\newblock [doi: 10.48550/arXiv.2304.02643].

\bibitem{dabov2007image}
K.~Dabov, A.~Foi, V.~Katkovnik, {\em et~al.}, ``Image denoising by sparse 3-d transform-domain collaborative filtering,'' {\em IEEE Transactions on Image Processing} {\bf 16}(8), 2080--2095  (2007).
\newblock [doi: 10.1109/TIP.2007.901238].

\bibitem{ravi2024sam2}
N.~Ravi, V.~Gabeur, Y.-T. Hu, {\em et~al.}, ``Sam 2: Segment anything in images and videos,'' {\em arXiv preprint arXiv:2408.00714}   (2024).
\newblock [doi: 10.48550/arXiv.2408.00714].

\bibitem{sengupta2024is}
S.~Sengupta, S.~Chakrabarty, and R.~Soni, ``Is sam 2 better than sam in medical image segmentation?,'' in {\em Medical Imaging 2025: Image Processing},  O.~Colliot and J.~Mitra, Eds.,  {\bf 13406}, International Society for Optics and Photonics, SPIE  (2025).
\newblock [doi: 10.1117/12.3047370].

\end{thebibliography}
\bibliographystyle{spiejour}   

\end{spacing}
\end{document}